\documentclass[11pt]{article}
\usepackage{curves}

\newcommand{\dslash}{\partial\hskip-6.28pt /}
\newcommand{\pa}{\partial}
\newcommand{\be}{\begin{equation}}
\newcommand{\ee}{\end{equation}}
\newcommand{\bea}{\begin{eqnarray}}
\newcommand{\eea}{\end{eqnarray}}
\newcommand{\tq}{\tilde{q}}
\newcommand{\hx}{\hat{x}}

\newcommand{\bt}[1]{{\bar t}}
\usepackage{amsmath,amsfonts,epsfig,color,latexsym}
\usepackage{amssymb}
\begin{document}
\input FEYNMAN

\thispagestyle{empty}
\null\vskip-24pt \hfill AEI-2002-070 \vskip 0pt \hfill  CERN-TH/2002-232
\vskip 0pt \hfill  LAPTH-935/02 \\
\vskip 0pt \hfill {\tt hep-th/0209103}
\vskip0.2truecm
\begin{center}
\vskip 0.2truecm {\Large\bf
A note on the perturbative properties of BPS operators
}\\
\vskip 1truecm
{\bf Gleb Arutyunov$^{*,**}$\footnote{email:{\tt 
agleb@aei-potsdam.mpg.de}},
Emery Sokatchev$^{\ddagger,\ddagger\ddagger}$ \footnote{email:{\tt Emery.Sokatchev@cern.ch} \\
$^{**}$On leave of absence from Steklov Mathematical Institute, Gubkin str.8,
117966, Moscow, Russia }
}\\
\vskip 0.4truecm
$^{*}$ {\it Max-Planck-Institut f\"ur Gravitationsphysik,
Albert-Einstein-Institut, \\
Am M\"uhlenberg 1, D-14476 Golm, Germany}\\
\vskip .2truecm $^{\ddagger}$ {\it CERN Theoretical Division, CH-1211 Geneva
23, Switzerland\\
$^{\ddagger\ddagger}$ Laboratoire d'Annecy-le-Vieux de Physique
Th\'{e}orique\footnote[3]{UMR 5108 associ{\'e}e {\`a}
 l'Universit{\'e} de Savoie} LAPTH\\ BP 110, F-74941 Annecy-le-Vieux Cedex, France
} \\
\end{center}

\vskip 1truecm \Large
\centerline{\bf Abstract}\vskip5mm \normalsize We discuss the 
perturbative behavior of the 1/2 BPS operators in ${\cal N}=2$ SCFT on 
the example of two very similar quadrilinear composite operators made 
out of hypermultiplets. An explicit one-loop computation shows that one 
of them is protected while the other acquires an anomalous dimension. 
Although both operators are superconformal primaries in the free case, 
the quantum corrections make the latter become a 1/2 BPS descendant of 
the Konishi multiplet, while the former remains primary. The comparative 
study of these two operators at higher orders may be helpful in 
understanding the quantum properties of the Konishi multiplet.

\newpage
\setcounter{page}{1}\setcounter{footnote}{0}
\section{Introduction}
A lot of progress in understanding the classical and quantum properties of the
superconformal field theories has been achieved since the discovery of the
AdS/CFT correspondence. This includes both the development of the abstract
superconformal representation theory as well as the analysis of the correlation
functions of concrete composite operators obtained in different regimes of the
correspondence.

A special class of composite operators in ${\cal N}=4$ SYM playing a 
privileged r\^ole in the AdS/CFT dictionary are the so-called BPS short 
operators, i.e. operators annihilated by a certain fraction of the 
Poincar\'e supercharges $Q$. They form isolated series of 
representations of the superconformal group with quantized (protected) 
conformal dimension (for a full classification see, e.g., \cite{dp}).

It is important to realize that the symmetry protects such BPS operators 
only if they are superconformal primary, i.e., if they are annihilated 
by all the generators $S$ of special conformal supersymmetry. For a 
given composite operator in ${\cal N}=4$ SYM the latter property is not 
obvious. It may happen that an operator looks BPS short (it is 
annihilated by some charges $Q$), but in fact it is a superconformal 
descendant of another, long multiplet. At the same time, such an 
operator may still be a conformal primary (i.e., it is annihilated by 
the conformal boosts) pure state and may have well-defined but 
unprotected conformal dimension. The issue is rather subtle because the 
superconformal ``primarity" is lost only at the quantum level. This 
possibility was first pointed out in \cite{Witten} in the ${\cal N}=1$ 
framework and has since been reiterated many times (see, e.g., 
\cite{AFSZ}).

The fact that a long supermultiplet lying on the threshold of the
unitary bound in the classical theory is a reducible representation of
the superconformal group has been known for a long time (see, e.g.,
\cite{dp}). Recently, a detailed analysis of the decomposition of long
multiplets into semishort or BPS short multiplets has been carried out
in \cite{DO}. In particular, when the superconformal primary long
multiplet of the ${\cal N}=4$ (${\cal N}=2$) algebra is a Lorentz
scalar, one finds 1/4 (1/2) BPS multiplets in its decomposition.

In this note we present a concrete and simple example of a ``fake" and a
``true" 1/2 BPS operators in ${\cal N}=2$ SCFT and discuss their nature 
in perturbation theory. After recalling the necessary tools to define 
the Konishi superfield in the framework of the ${\cal N}=2$ superspace 
in Section 2, in Section 3 we extract from it a superconformal 
descendant (but a conformal primary) having the quantum numbers and the 
appearance of an ${\cal N}=2$ 1/2 BPS operator. It is obtained by acting 
on the Konishi superfield with four (i.e., half of the total number) 
spinor derivatives and by using the interacting field equations for the 
elementary constituents. This operator is a single-trace quadrilinear 
combination of the elementary hypermultiplet (HM) superfields and 
involves color group commutators.\footnote{It was precisely the presence 
of a commutator term in the definition of a composite chiral operator 
that was interpreted in \cite{Witten} as a signal that it may be a 
descendant of some lower dimensional non-chiral operator. For relatively 
low conformal dimensions this criterium for identifying descendant 
operators in perturbation theory was used, e.g. in \cite{BKRS,Ryzhov}.} 
At the same time, we construct another, very similar composite with 
fully symmetrized color indices which is supposed to be a true protected 
1/2 BPS primary operator. Our present understanding of the short 
composite operators does not allow us to make definitive statements 
about the latter, we can only rely on perturbative evidence 
\cite{tests}. In Section 4 we examine the one-loop corrections to the 
two-point functions of both operators and show explicitly the origin of 
the anomalous dimension of the former. Since this is precisely the 
anomalous dimension of the ``parent" Konishi multiplet, understanding 
the mechanism which distinguishes the ``fake" 1/2 BPS operator from the 
very similar ``true" one at higher orders in perturbation theory may 
prove useful in an attempt to compute the exact anomalous dimension of 
the Konishi multiplet. In addition, the example that we consider in the 
paper can serve as a concrete and simple illustration of the general 
representation theory discussion in Ref. \cite{DO}. Finally, in Section 
5 we formulate several open questions, the main one being: How to test 
the ``primarity" of a given operator?

\section{${\cal N}=4$ SYM in terms of ${\cal N}=2$ superfields}

The two ${\cal N}=2$ ingredients of the ${\cal N}=4$ SYM theory are the
HM and the ${\cal N}=2$ SYM multiplet. In harmonic superspace
\cite{GIOS} the former is described by a Grassmann
(G-)analytic\footnote{Grassmann analyticity is the natural
generalization of the notion of chirality in the case of extended
supersymmetry.} (also frequently called ``1/2 BPS") superfield of $U(1)$
charge $+1$,
\begin{equation}\label{defq}
  D^+_\alpha q^+ = \bar D^+_{\dot\alpha} q^+ = 0 \ \Rightarrow \ q^+ =
  q^+(x,\theta^+,\bar\theta^+,u)\,.
\end{equation}
Here $D^+_{\alpha,\dot\alpha} = u^+_i D^i_{\alpha,\dot\alpha}$ and $u^\pm_i$,
$i=1,2$ form an $SU(2)$ matrix of harmonic variables on the two-sphere $S^2
\sim SU(2)/U(1)$. In the free case the HM satisfies the equation of motion
\begin{equation}\label{eqm}
  D^{++}q^+ = 0\;,
\end{equation}
which takes the form of a harmonic (H-)analyticity condition. Here
\begin{equation}\label{hmder}
  D^{++} = \pa^{++}_u - 2i\theta^+\sigma^\mu\bar\theta^+ \pa_\mu
\end{equation}
is the supercovariant harmonic derivative in the G-analytic basis, which also
plays the r\^ole of the raising operator of $SU(2)$. The solution of eq.
(\ref{eqm}) is the ultrashort on-shell superfield
\begin{equation}\label{onshsf}
  q^+ = q^i(x)u^+_i +\theta^{+\alpha}\psi_\alpha(x)
+\bar\theta^+_{\dot\alpha}\bar\kappa^{\dot\alpha}(x)
+2i\theta^+\sigma^\mu\bar\theta^+ \partial_\mu f^i(x)u^-_i\,,
\end{equation}
where the physical components of the ${\cal N}=2$ HM, the doublet of
scalar fields $q^i(x)$ and the fermions $\psi,\ \bar\kappa$, satisfy
their free field equations
$$ \square q^i(x)=\dslash\psi
=\dslash\bar\kappa=0\,.
$$

The ${\cal N}=2$ SYM multiplet is described by another G-analytic superfield,
this time of $U(1)$ charge $+2$, $V^{++}(x,\theta^+,\bar\theta^+,u)$. It plays
the r\^ole of the gauge connection in the covariantized harmonic derivative
\begin{equation}\label{hdcov}
  \nabla^{++} = D^{++} + ig V^{++}\;,
\end{equation}
so that the flat commutation relations with $D^+_{\alpha,\dot\alpha}$  are
preserved:
\begin{equation}\label{flatcom}
  [\nabla^{++}, D^+_{\alpha,\dot\alpha}] = 0\,.
\end{equation}
In the Wess-Zumino gauge the component content of $V^{++}$ is reduced to the
off-shell ${\cal N}=2$ gauge multiplet:
\begin{eqnarray}\label{6.10.1}
 V^{++}_{\rm WZ} &=& -2i\theta^+\sigma^\mu\bar\theta^+
A_\mu(x) - i \sqrt2 (\theta^+)^2\bar \phi(x) + i\sqrt2(\bar\theta^+)^2 \phi(x)
\nonumber \\ && +\,4(\bar\theta^+)^2\theta^{+\alpha}\psi^i_{\alpha}(x) u^{-}_i
-4(\theta^+)^2\bar\theta^+_{\dot\alpha}\bar\psi^{\dot\alpha i }(x) u^{-}_i
\nonumber \\ && +\, 3(\theta^+)^2(\bar\theta^+)^2 D^{ij}(x)u^-_iu^-_j\,.
\end{eqnarray}
In the ${\cal N}=4$ theory the HM $q^+$ belongs to the adjoint representation
of the gauge group and interacts with the gauge multiplet through the usual
minimal coupling:
\begin{equation}\label{6}
  \nabla^{++} q^+ = D^{++} q^+ + ig\  [V^{++},q^+] = 0\,.
\end{equation}

The harmonic gauge connection $V^{++}$ serves as the prepotential of the ${\cal
N}=2$ SYM theory. The derivatives $D^+_{\alpha,\dot\alpha}$ do not need a
connection in the frame where G-analyticity is manifest (see (\ref{flatcom})).
The connections for the conjugate harmonic derivative $\nabla^{--} = D^{--} +
ig V^{--}$ and for the spinor derivatives $\nabla^-_{\alpha,\dot\alpha} =
D^-_{\alpha,\dot\alpha} + ig A^-_{\alpha,\dot\alpha}$ are determined through
the conventional constraints
\begin{eqnarray}
  &&[\nabla^{++},\nabla^{--}] = 2D^0\,, \label{concon'}\\
  &&[\nabla^{--},D^+_{\alpha,\dot\alpha}] = \nabla^-_{\alpha,\dot\alpha} \,,
   \label{concon}
\end{eqnarray}
where $D^0$ is the $U(1)$ charge operator. Next, the field strength is defined
by the relation
\begin{equation}\label{1}
  \{\nabla^+_\alpha, \nabla^-_\beta\} = 2ig\ \epsilon_{\alpha\beta}\bar W\,.
\end{equation}
It  is covariantly chiral,
\begin{equation}\label{2}
  \nabla^+_\alpha\bar W   = \nabla^-_\alpha\bar W  = 0\,,
\end{equation}
and harmonic independent,
\begin{equation}\label{3}
  \nabla^{++} \bar W   = 0\,,
\end{equation}
as a consequence of the above constraints.

In the free case the curvature $\bar W$ satisfies the field equation
\begin{equation}\label{4}
  \bar D^+_{\dot\alpha} \bar D^{+\dot\alpha}  \bar W   = 0\,.
\end{equation}
The coupling to ${\cal N}=2$ HM matter modifies this equation into
\begin{equation}\label{5}
  \bar D^+_{\dot\alpha} \bar D^{+\dot\alpha} \bar W   = 8ig\ [\tilde q^+,
  q^+]\,,
\end{equation}
where $\tilde q^+$ denotes a special conjugation on $S^2$ which preserves
G-analyticity.

Eqs. (\ref{6}) and (\ref{5}) are equivalent to the ${\cal N}=4$ SYM field
equations written down in terms of ${\cal N}=2$ superfields. They can be
obtained from an off-shell action with manifest ${\cal N}=2$ supersymmetry,
which allows a straightforward quantization.


\section{The Konishi multiplet and its 1/2 BPS descendant}

The ${\cal N}=2$ Konishi superfield is defined as follows:
\begin{equation}\label{7}
  K = {\rm Tr} \left(\frac{1}{2}\bar W W + \tilde q^+ \nabla^{--} q^+ -
   q^+ \nabla^{--}\tilde q^+  \right) =
   {\rm Tr} \left( \bar\phi \phi + \bar q^i q_i \right) + \theta\ {\rm terms}\
   .
\end{equation}
Using the constraints (\ref{defq}), (\ref{concon}), (\ref{1}), (\ref{2}) and
the field equations (\ref{6}), (\ref{5}) it is easy to check that
\begin{equation}\label{9}
  D^{+\alpha}D^+_\alpha K = 12ig\ {\rm Tr} \left(\bar W [\tilde q^+, q^+]
  \right)
\end{equation}
and
\begin{equation}\label{9'}
  D^{++} K=0\,.
\end{equation}
Equation (\ref{9}) encodes the classical non-conservation of the Konishi
current $K= \ldots + \theta^i\sigma^\mu\bar\theta_i k_\mu(x) + \ldots $. The
H-analyticity condition (\ref{9'}) means that the chargeless superfield $K$ is
harmonic independent.\footnote{ Note that the similar operator $J = {\rm Tr}
\left(-\bar W W + \tilde q^+ \nabla^{--} q^+ - q^+ \nabla^{--}\tilde q^+
\right)$ corresponds to one of the ${\cal N}=2$ supercurrent multiplets and
satisfies the conservation conditions $D^{+\alpha}D^+_\alpha J = D^{++} J = 0$
even in the presence of interaction.}

The quantum corrections lead to an anomaly \cite{Konishi} in the right-hand
side of eq. (\ref{9}):
\begin{equation}\label{10}
  D^{+\alpha}D^+_\alpha K = 12ig\ {\rm Tr} \left(\bar W [\tilde q^+, q^+]
  \right) + cg^2\ \bar D^+_{\dot\alpha} \bar D^{+\dot\alpha}{\rm Tr} \left(\bar
  W\bar W \right)\,,
\end{equation}
where $c$ is a number. We can simplify eq. (\ref{10}) by hitting it with $\bar
D^+_{\dot\alpha} \bar D^{+\dot\alpha}$. The quantum anomaly term drops out; in
the classical term the derivatives see only $\bar W$ and we can use the field
equation (\ref{5}). The result is
\begin{eqnarray}
  (D^+)^4 K \equiv \bar D^+_{\dot\alpha} \bar D^{+\dot\alpha} D^{+\alpha}D^+_\alpha K
  &=& -96g^2\ {\rm Tr} \left([\tilde q^+, q^+] [\tilde q^+, q^+]\right)\nonumber\\
  &=& -192 g^2\left\{ {\rm Tr} \left(\tilde q^+ q^+\tilde q^+q^+\right) -
  {\rm Tr} \left(\tilde q^+\tilde q^+q^+ q^+\right)\right\} \,. \label{12}
\end{eqnarray}
The left-hand side of this equations is annihilated by all of the
G-analytic derivatives $D^+,\bar D^+$, and so is the right-hand side,
due to the G-analyticity of the HMs. Let us define the single-trace
composite operator of $U(1)$ charge $+4$ (equal to its naive dimension)
\begin{equation}\label{kondes}
  {\cal K}^{+4} = {\rm Tr} \left([\tilde q^+, q^+] [\tilde q^+, q^+]\right)\,.
\end{equation}
It is manifestly G-analytic and in addition is also H-analytic (see eq.
(\ref{hdcov})):
\begin{equation}\label{hankon}
 D^+_\alpha {\cal K}^{+4} = \bar D^+_{\dot\alpha}{\cal K}^{+4} = D^{++}{\cal K}^{+4} = 0\,.
\end{equation}
Note the appearance of commutators under the trace in (\ref{kondes}), which
indicates that the operator has been derived through the use of the filed
equation (\ref{5}).

A very similar operator, also both G- and H-analytic, is obtained by taking the
sum in the second line of eq. (\ref{12}):\footnote{There exist two other
quadrilinear G- and H-analytic HM composites, ${\rm Tr} \left(\tilde q^+ q^+
q^+q^+\right)$ and ${\rm Tr} \left(q^+ q^+ q^+q^+\right)$, but they are not
related to the Konishi multiplet, so we do not consider them here. }
\begin{equation}\label{prot}
  {\cal O}^{+4} = {\rm Tr} \left(\tilde q^+ q^+\tilde q^+q^+\right) +
  {\rm Tr} \left(\tilde q^+\tilde q^+q^+ q^+\right) \,.
\end{equation}
Despite the appearance, the two operators ${\cal K}^{+4}$ and ${\cal O}^{+4}$
are substantially different, but this can only be seen in the interacting
quantum theory. In the free case both operators can be viewed as 1/2 BPS
operators. In fact, the free Konishi superfield $K_0$ of canonical dimension
two is a reducible representation of conformal supersymmetry (much like the
Konishi current $k^\mu(x)$ of canonical dimension three which is a reducible
representation of the conformal group). As such, it can be decomposed into
several irreducible multiplets (see \cite{DO} for details), one of which is the
1/2 BPS projection $ (D^+)^4 K _0$. Then the free version of eq. (\ref{12})
simply means that this 1/2 BPS projection vanishes. This statement is
manifestly super-Poincar\'e invariant because $[Q,(D^+)^4]=0$, but it is also
superconformal, since $[S, (D^+)^4]K_0 =0$ {\it provided that the superfield
$K_0$ has canonical dimension two.} Then, what remains in the free Konishi
superfield $K_0$ forms a {\it semishort multiplet}. As to the right-hand side
of eq. (\ref{12}), in the free case it is not related to the Konishi multiplet.
Still, as long as it is made out of free HMs, ${\cal K}^{+4}$ undoubtedly is a
superconformal primary 1/2 BPS multiplet, just like the other composite ${\cal
O}^{+4}$.

Now, it is well known that in the quantum interacting theory the Konishi
superfield $K$ acquires an anomalous dimension \cite{Anselmi,BKRS,AFP2}. As
soon as this happens, the left-hand side of eq. (\ref{12}) ceased to be
superconformally covariant, since $[S, (D^+)^4 ] K \neq 0$ if ${\rm dim} \ K >
2$. In this sense the 1/2 BPS projection $ (D^+)^4  K $ becomes a {\it
superconformal descendant} of the now {\it long} Konishi multiplet $K$.
Nevertheless, this 1/2 BPS projection is still {\it conformal primary} in the
following sense: It can be checked that although  the differential operator
$(D^+)^4$ does not commute with the conformal boost generators $K^\mu$,
$[K^\mu,  (D^+)^4 ] \propto \bar D^+\sigma^\mu D^+\; D^{++}$, when applied to
the Konishi superfield this commutator still vanishes because of the
H-analyticity (or $SU(2)$ irreducibility) condition (\ref{9'}). The above
discussion applies to the quadrilinear composite operator ${\cal K}^{+4}$ as
well, since it is related to the 1/2 BPS {\it non-covariant} projection of $K$
through the field equation (\ref{12}). We conclude that in the interacting {\it
quantum} theory ${\cal K}^{+4}$ should be regarded as a superconformal
descendant (but at the same time conformal primary) of the long Konishi
multiplet.

On the other hand, there is no obvious reason why the second 1/2 BPS operator
${\cal O}^{+4}$ should change its nature when the interaction is switched on.
It does not seem to be related via the field equations to any other, long
multiplet, so it is likely to remain superconformal primary. If this is true,
its dimension should be protected by superconformal symmetry. However, our
present understanding of the making of composite operators does not give us a
direct and simple test for ``primarity". The (non)preservation of this
property in the quantum theory is somehow related to the potentially singular
nature of the composite operators. It is an interesting problem to find out
what exactly ``goes wrong" with ${\cal K}^{+4}$ but not with ${\cal O}^{+4}$.
The best we can do at present is to make perturbative tests, see the next
section.

Note that in a similar manner we can construct 1/2 BPS operators of
lower dimension (charge). These are bilinear or trilinear in the
elementary HMs constituents:
\begin{equation}\label{bitri}
  {\cal O}^{+2}_1 = {\rm Tr} \left(q^+ q^+\right) , \quad
  {\cal O}^{+2}_2 = {\rm Tr} \left(\tilde q^+ q^+\right) , \quad
  {\cal O}^{+3}_1 = {\rm Tr} \left( q^+ q^+ q^+\right) , \quad
  {\cal O}^{+3}_2 = {\rm Tr} \left(\tilde q^+ q^+ q^+\right)
\end{equation}
and conjugates. At this low dimension there can be no commutators under the
trace, therefore we can expect that the above operators are primary and thus
protected. In fact, the bilinears ${\cal O}^{+2}$ are current multiplets, they
contain various components of the R symmetry current of the ${\cal N}=4$
theory. This gives an additional reason why they should be protected.

Another remark concerns the clear difference in the structure of the descendant
${\cal K}^{+4}$ and its ``parent" superfield $K$ (\ref{7}). The latter is
bilinear in the matter superfields $q^+$, but non-polynomial in the gauge
superfield.\footnote{ In the harmonic formulation the non-linear object is the
connection $V^{--}$ whose expression in terms of the prepotential $V^{++}$ via
the constraint (\ref{concon'}) is non-polynomial \cite{GIOS}; in the more
familiar ${\cal N}=1$ formulation the non-polynomial dependence is due to the
gauge factor in ${\rm Tr} \; \bar\Phi e^{igV} \Phi$.} However, its descendant
${\cal K}^{+4}$ is polynomial in the HMs and does not involve any gauge
superfields. In this respect it strongly resembles the (supposedly) primary
operator ${\cal O}^{+4}$ (\ref{prot}). We believe that the much simpler
structure of ${\cal K}^{+4}$ makes it a better candidate for further
investigations than its parent $K$.


\section{BPS operators in perturbation theory}
As discussed in the preceding section, the quadrilinear composites
${\cal K}^{+4}$ and ${\cal O}^{+4}$, although having the same properties
of 1/2 BPS operators in the free case, are expected to behave quite
differently in the (perturbative) quantum theory. To get a better
feeling of how this happens, it is useful to reexamine the two-point
functions of these operators at one loop. Below we outline this
calculation for ${\cal K}^{+4}$ in the ${\cal N}=2$ harmonic superspace
formalism \cite{GIOS}.

To start with, Wick contractions with the HM Euclidean propagator $\langle
\tq^+(1)q^+(2) \rangle_{\theta=0}={(12)}/{4\pi^2 x^2_{12}}$ produce the free
two-point function\footnote{The color combinatorics is performed by using the
relations $\mbox{Tr}(T^aT^b)=\delta^{ab}$, $f_{acd}f_{bcd}=2N\delta_{ab}$,
$f_{amn}f_{bnp}f_{cpm}=Nf_{abc}$, where $T^a$ with $a=1,\ldots N^2-1$ are the
$SU(N)$ Lie algebra generators.} \bea \langle {\cal K}^{+4}(x_1){\cal
K}^{+4}(x_2) \rangle_{\theta=0}=
\frac{12N^3(N^2-1)}{(4\pi^2)^4}\frac{(12)^4}{x_{12}^8} \, ,\eea where
$(12)\equiv u^{+i}_1u_{2i}^+$ is the $SU(2)$ invariant contraction of the
harmonic variables at the two points. Here and below we consider only the
lowest component of ${\cal K}^{+4}$ putting thereby $\theta_1=\theta_2=0$.

At order $g^2$ we have graphs of two types, $J_1$ and $J_2$ (see Fig.
1).\footnote{The self-energy contribution to the HM propagator vanishes
in the ${\cal N}=2$ formalism.} The graphs of the type $J_1$ are
obtained by inserting a gluon propagator between any two (solid) HM
lines. The graphs of the type $J_2$ always contain two HM loops
connected by a gluon exchange.

\begin{center}
\begin{picture}(10000,20000)(14000,-10000)
\drawline\gluon[\N\CENTRAL](5000,0)[2]
\global\Ytwo = \gluonbacky
\global\Yone= \gluonbacky
\global\Ythree= \gluonbacky
\global\Yfour= \gluonbacky
\multroothalf\Yone \multroothalf\Yone
\curve(0,\Yone,5000,\Ytwo,10000,\Yone)
\drawarrow[\E\ATTIP](4700,\Ytwo)
\curve(0,\Yone,5000,0,10000,\Yone)
\drawarrow[\W\ATTIP](4200,0)
\global\advance\Ythree by 20
\global\advance\Yfour by 2000
\curve(0,\Yone,5000,\Yfour,10000,\Yone)
\drawarrow[\E\ATTIP](5000,\Yfour)
\curve(0,\Yone,5000,-2000,10000,\Yone)
\drawarrow[\W\ATTIP](4500,-2000)
\put(4600,-4000){$J_1$}

\curve(15000,\Yone,20000,\Ytwo,25000,\Yone)
\drawarrow[\E\ATTIP](20000,\Ytwo)
\curve(15000,\Yone,20000,0,25000,\Yone)
\drawarrow[\W\ATTIP](19700,0)
\drawline\gluon[\E\CENTRAL](25000,\Yone)[4]
\global\Xone=\gluonbackx
\global\Xtwo=\particlemidx
\curve(\Xone,\Yone,35000,\Ytwo,40000,\Yone)
\drawarrow[\E\ATTIP](35000,\Ytwo)
\curve(\Xone,\Yone,35000,0,40000,\Yone)
\drawarrow[\W\ATTIP](34700,0)
\curve(15000,\Yone,\Xtwo,\Yfour,40000,\Yone)
\drawarrow[\E\ATTIP](\Xtwo,\Yfour)
\curve(15000,\Yone,\Xtwo,-2000,40000,\Yone)
\drawarrow[\W\ATTIP](\Xtwo,-2000)
\put(\Xtwo,-4000){$J_2$}

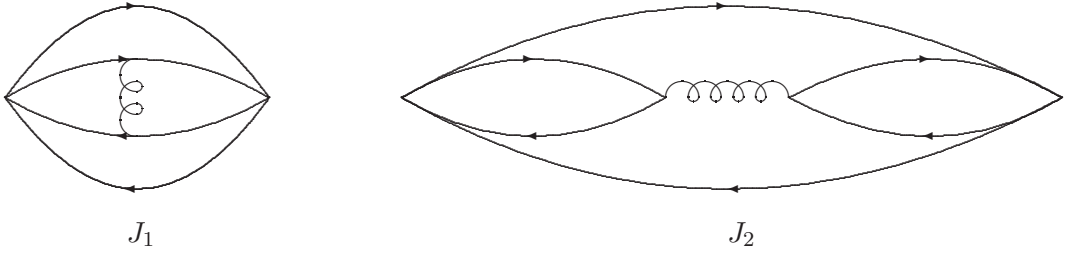
\put(2000,-6000){ Figure 1. Two different topologies contributing to the
two-point function of ${\cal K}^{+4}$.} \put(2000,-7200){ Only the
second graph is responsible for the anomalous dimension.}

\end{picture}
\end{center}

Explicitly, $J_1$ and $J_2$ are given by the following analytic superspace
integrals: \bea J_1\sim \frac{(12)^2}{x^4_{12}}\ \int
du_{5,6}d^4\theta_{5,6}^+d^4x_{5,6}~ \frac{(15)(52)}{\hx_{15}^2\hx_{52}^2}\
\frac{(16)(62)}{\hx_{16}^2\hx_{62}^2} \langle V^{++}(5)V^{++}(6)\rangle \, \eea
and \bea J_2\sim \frac{(12)^2}{x^4_{12}}\ \int
du_{5,6}d^4\theta_{5,6}^+d^4x_{5,6}~ \frac{(15)(51)}{\hx_{15}^2\hx_{51}^2}\
\frac{(26)(62)}{\hx_{26}^2\hx_{62}^2} \langle V^{++}(5)V^{++}(6)\rangle \, .
\eea
 Here $$\langle V^{++}(5)V^{++}(6)\rangle =\frac{(\theta_5^+-(56^-)\theta_6^+)^4
}{2\pi^2x_{56}^2} \delta^{(-2,2)}(u_5,u_6)$$ is the propagator of the ${\cal
N}=2$ prepotential $V^{++}$ in the Feynman gauge. With all external $\theta$'s
set to zero, the superinvariant coordinate difference entering the HM
propagator between points, e.g. 1 and 5 is defined as follows: \bea
\hx^2_{15}=\left(x_{15}-2i\frac{(15^-)}{(15)}\theta^+_5\sigma\bar{\theta^+}_5\right)^2
=\hx^2_{51} \, ,\eea where $(15^-) = u^{+i}_1 u^-_{5i}$. The integration over
the internal Grassmann and harmonic variables is very easy and we find for
$J_1$ and $J_2$ the following space-time expressions (dropping the propagator
factors): \bea J_1 \sim\Box_1 \int \frac{dx_5dx_6 }{ x_{15}^2 x_{16}^2 x_{56}^2
x_{25}^2x_{26}^2 }\, , ~~~~~~ J_2\sim \int\frac{dx_5} {x_{15}^4 x_{25}^4} \, .
\eea The integral in $J_1$ is finite, therefore conformal invariance tells us
that it is proportional to $1/x_{12}^2$. Thus $J_1$ contributes only a contact
term to the two-point function which vanishes for separated points. The
integral $J_2$ is logarithmically divergent and computing it, for instance in
dimensional regularization, one finds a simple pole, $J_2\sim 1/\epsilon$. The
residue determines the anomalous dimension. Performing a multiplicative
renormalization and doing carefully the color combinatorics, we find for the
renormalized operator ${\cal K}^{+4}$ the one-loop anomalous dimension
$\gamma={3g^2N}/{4\pi^2}$, i.e. precisely the same as for the renormalized
Konishi operator.

This result is in agreement with the discussion in Section 3, were
${\cal K}^{+4}$ was shown to arise as a superconformal descendant of the
Konishi multiplet. We clearly see that the 1/2 BPS operator ${\cal
K}^{+4}$, which was a superconformal primary in the free theory, {looses 
this property} when the quantum corrections are turned on. Had it 
remained primary, its dimension would have been protected by the 
superconformal symmetry.

In fact, this is what happens to the operator ${\cal O}^{+4}$. Repeating our
one-loop calculation for the two-point function of ${\cal O}^{+4}$, this time
we find that only the graph $J_1$ contributes while $J_2$ is absent due to the
mismatch between the color antisymmetrization at the interaction vertices and
the symmetrization at the external points. However, as we have just seen, the
contribution of $J_1$ is a contact term, so the two-point function of ${\cal
O}^{+4}$ does not receive any quantum correction at one loop (for
non-coincident points).\footnote{This result is stronger than just saying that
the operator ${\cal O}^{+4}$ has a protected dimension. In fact, this is what
is believed to happen to all 1/2 BPS primary operators \cite{3pts}.} 

Finally, applying the same arguments to the mixed two-point function 
$\langle {\cal K}^{+4} {\cal O}^{+4}\rangle$, we again find that the 
graph $J_2$ cannot contribute, whereas $J_1$ still gives a contact term. 
Thus, the two operators do not mix, at least at one loop. This situation 
is different from the example in Ref. \cite{Ryzhov} concerning two 
dimension four operators in the $\underline{84}$ of $SU(4)$. One of them 
is a single-trace descendant, but the other is double-trace and they mix 
at one loop.

The above perturbative test gives indirect evidence that the operator
${\cal O}^{+4}$ remains superconformal primary in the quantum theory,
while ${\cal K}^{+4}$ changes its nature. It would be preferable to have 
another, direct way of testing ``primarity".

Clearly, the descendant ${\cal K}^{+4}$ has the same anomalous dimension as its
parent operator. Therefore, from the practical point of view, it might be
easier to study the quantum properties of $K$ by instead examining ${\cal
K}^{+4}$. Indeed, ${\cal K}^{+4}$ is an analytic gauge-invariant composite
which, contrary to $K$, does not involve any covariantizing gauge connection in
its definition. Moreover, the existence of a protected superconformal operator
with the same quantum numbers and with a very similar structure suggests
important simplifications. It is conceivable that those graphs which appear in
the perturbative expansion of the two-point functions for both ${\cal O}^{+4}$
and ${\cal K}^{+4}$ should have a vanishing (or rather, contact term)
contribution, as we have just seen in the simplest one-loop example. If this is
true, a large subset of graphs could be eliminated from the calculation of the
anomalous dimension of ${\cal K}^{+4}$. This point, however, requires further
study because at higher loops the protection mechanism for ${\cal O}^{+4}$
might be more involved, perhaps not due to the convergence of individual graphs
but rather to non-trivial color combinatorics leading to cancellation of
individual divergences. In any case, the question of understanding how an
operator like ${\cal O}^{+4}$ might be protected to all orders in perturbation
theory is still open and certainly very interesting.

\section{Conclusions}
In this note, in the context of ${\cal N}=2$ SCFT we have demonstrated the
different perturbative behavior of operators having the same quantum numbers of
a 1/2 BPS state. We considered the simplest example of two analytic superfields
composed of four HMs. Both of them are superconformal primary operators in the
free theory but when the quantum effects are turned on, the first one ceased to
be primary with respect to the special conformal supersymmetry and becomes a
member of the long unprotected Konishi multiplet. The second one remains a
superconformal primary at the quantum level and hence is protected.

As argued in Ref. \cite{DO}, in some special cases one is able to tell 
that a short or a semishort operator with given quantum numbers cannot 
possibly be contained as a descendant in another, long multiplet, and 
should thus be automatically protected. Our examples are clearly not of 
this type. At present, we do not have a better way to decide whether one 
or the other of our 1/2 BPS operators could acquire an anomalous 
dimension but making perturbative tests. In this context we remark that 
the various ``non-renormalization theorems" for short operators in the 
literature have always been based on the (implicit) assumption of 
``primarity". As we have tried to show in this note, demonstrating the 
``primarity" of a given composite short operator at the quantum level 
seems equally difficult, in practice equivalent to proving that it is 
protected. It would be highly desirable to find some intrinsic criterion 
for determining the superconformal properties of an operator in 
perturbation theory.

The discussion above raises the intriguing question about what precisely 
is responsible for the loss of ``primarity" of certain operators. 
Progress in this direction might lead to explaining the mechanism which 
protects short operators, on the one hand. On the other, it may prove 
helpful in revealing the deep origin of the anomalous dimension of long 
multiplets and eventually to its exact evaluation.

Our discussion also applies to the so-called BMN operators recently 
introduced in the context of the SYM/plane-wave correspondence 
\cite{BMN}. Indeed, a BMN operator with impurities is a mixture of 
operators in different $SU(4)$ representations. It contains, in 
particular, irreducible $SU(4)$ pieces which are single-trace operators 
involving commutators of elementary fields. Thus, some of them might be 
superconformal descendants of lower-dimensional operators. Since the 
span of dimensions in a supermultiplet is finite, it is conceivable that 
in the BMN limit of infinite quantum numbers the difference between a 
primary field and its supersymmetry descendant actually disappears or 
becomes insignificant. It would be interesting to better understand this 
point.

\section*{Acknowledgements} We are grateful to  C. Callan,
S. {}Ferrara, P. Howe, H. Osborn and E. Sezgin for many useful
discussion. G. A. was supported by the DFG and by the European
Commission RTN programme HPRN-CT-2000-00131, and in part by RFBI grant
N99-01-00166.

\newpage

\renewcommand{\baselinestretch}{0.6}

\end{document}